%% file: longmore_s_g305.tex
\documentclass[useAMS,usenatbib,psfig]{mn2e}
\usepackage{graphicx,lscape,multirow}
\usepackage{natbib}
\newcommand{\hii}{{H{\scriptsize II} }}

\title[Embedded Stars toward Massive Star Forming Regions]{Embedded Stellar Populations towards Young Massive Star Formation Regions I. G305.2+0.2}

\author[S.N.Longmore et al.]{S. N. Longmore$^{1,2}$\thanks{E-mail:
snl@phys.unsw.edu.au}, M. Maercker$^{3}$, S. Ramstedt$^{3}$, M.G. Burton$^{1,4}$ \\ 
$^{1}$School of Physics, University of New South Wales, Kensington, 2052, Sydney, Australia\\ 
$^{2}$Australia Telescope National Facility CSIRO, Epping, Sydney, NSW 1710, Australia\\ 
$^{3}$Stockholm Observatory, AlbaNova University Center, 106 91 Stockholm, Sweden\\
$^{4}$Armagh Observatory, College Hill, Armagh, BT61 9DG }

\begin{document}

\date{ }

\pagerange{\pageref{firstpage}--\pageref{lastpage}} \pubyear{2006}

\maketitle

\begin{abstract}
We present deep, wide-field J, H and Ks images taken with IRIS2 on the
Anglo Australian Telescope, towards the massive star formation region
G305.2+0.2. Combined with 3.6, 4.5, 5.8 and 8.0$\mu$m data from the
GLIMPSE survey on the Spitzer Space Telescope, we investigate the
properties of the embedded stellar populations.  After removing
contamination from foreground stars we separate the sources based on
their IR colour. Strong extended emission in the GLIMPSE images
hampers investigation of the most embedded sources towards the known
sites of massive star formation. However, we find a sizeable
population of IR excess sources in the surrounding region free from
these completeness effects. Investigation reveals the recent star
formation activity in the region is more widespread than previously
known.

Stellar density plots show the embedded cluster in the region,
G305.24+0.204, is offset from the dust emission. We discuss the effect
of this cluster on the surrounding area and argue it may have played a
role in triggering sites of star formation within the region. Finally,
we investigate the distribution of IR excess sources towards the
cluster, in particular their apparent lack towards the centre compared
with its immediate environs. 
\end{abstract}


\section{Introduction}
Reprocessing of stellar light by circumstellar material around young
stars is observed at infrared (IR) wavelengths as excess emission
above that of a blackbody. The extent of the IR excess provides a
measure of the amount of circumstellar material and hence an
indication of the evolutionary state of the object
\citep[e.g.][]{adams1987}. Traditionally, the J, H and K bands (1.2,
1.6 and 2.2 $\mu$m) have successfully been used to identify IR excess
sources \citep[e.g.][]{ladaadams1992} but recent work including L-band
(3.6$\mu$m) data has highlighted that IR-excess sources are much more
clearly separated using longer wavelengths \citep{kenyonhartmann1995,
maerckerburton2005, maercker2006}. With large optical depths of
material towards the youngest regions, mid-IR observations at high
enough spatial resolution to resolve individual sources are required
to uncover the most heavily embedded objects
\citep[e.g.][]{longmore2006}. By combining deep, wide-field, near-IR
images of selected regions with catalogued mid-IR data, we aim to
identify IR-excess sources, uncover their spatial distribution and
investigate the star formation history of the regions. For this first
paper, we have chosen a region which has also been studied on the same
telescope with near-IR spectroscopy of selected objects
\citep{leistra2005}. Using these, we can calibrate our derived
photometric spectral types and so compare results from the automated
matching routines to check their accuracy.

\subsection{The G305 complex}
\begin{figure}
  \begin{center}
    \includegraphics[width=0.45\textwidth, trim=0 0 -5 0]{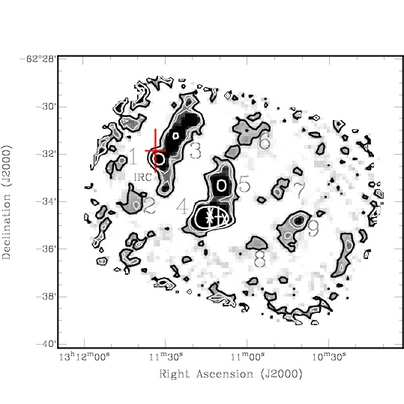}
    \caption{1.2mm dust continuum map of the G305.20+0.2 region
    \citep[from][]{hill2005} overlayed with black contours at
    5$\sigma$ and white contours at 10, 15 and 50$\sigma$ levels. The
    letters `IRC' show the position of the infrared cluster,
    G305.24+0.204. Plus and cross symbols show the position of
    methanol and water masers respectively. Regions of emission are
    numbered 1 to 9 and designated MM1, MM2 etc. for future
    reference. The apparent emission at the edge of the image are
    artefacts from the mapping process.}
    \label{fig:g305_simba}
  \end{center}
\end{figure}

The $\sim 1.5 ^\circ \times 1.5^\circ$ region at l $\sim$ 305$^\circ$
within the Scutum-Crux arm of the galaxy, designated G305 by
\citet{georgelin1988}, is a large ($\sim$ 100pc in extent) complex of
recent and active star formation at a distance of between 3.3 and 4kpc
\citep{leistra2005}.  With an age of 3-5 Myr, the complex is one of
the most luminous giant HII regions in the galaxy containing in excess
of 31 O stars. \citet{clarkporter2004} have recently summarised the
global properties of the complex which include two optically revealed
clusters, several embedded IR clusters, significant HII emission, many
UCHII regions and sites of maser emission (see \citet{clarkporter2004}
for references and further details). Evidence suggests that the
optically revealed clusters (Danks 1 \& 2) and associated Wolf-Rayet
star (WR 48a) are driving a wind-blown bubble surrounding the entire
region, which is responsible for triggering a second generation of
star formation -- the embedded near-IR clusters -- towards the edge of
the bubble. This includes the cluster G305.24+0.204, on the NW edge,
which contains several late O/early B stars \citep{leistra2005} that
are only apparent at IR wavelengths. Figure~\ref{fig:g305_simba} shows
the 1.2mm continuum emission \citep[from][]{hill2005} in the vicinity
of this cluster (the position of which is labelled as `IRC') and shows
several dust emission regions.
The brightest 1.2mm core (MM4 in Figure~\ref{fig:g305_simba}) has
previously been studied from the near-IR \citep{debuizer2003} through
mid-IR \citep{walsh2001}, mm continuum \citep{hill2005}, mm molecular
line \citep{walsh_burton2006}, 6.67GHz Class II methanol maser
emission \citep{norris1993} and cm continuum
\citep{phillips1998,walsh2002}.  These observations reveal both
ultra-compact HII regions and even younger massive stellar sources
within the core, suggesting it harbours a deeply embedded massive
protocluster. Furthermore, molecular line observations near MM5 and 8
\citep[G305N and G305SW respectively from][]{walsh_burton2006} show
the molecular gas to be both cold and quiescent, and as such may be
examples of pre-stellar cores. With multiple epochs of star formation
in the region, we aim to investigate the recent star formation history
by identifying the embedded and youngest sources through their IR
colours.

\section{IRIS2 Observations and Data Reduction}
\label{sec:iris2_obs}
The J, H and Ks-band (1.2, 1.6 and 2.2$\mu$m) images were observed
using IRIS2\footnote{IRIS2 employs a 1024x1024 Rockwell HAWAII-1
HgCdTe infrared detector with a plate scale of 0.4486 $\pm$
0.0002$\arcsec$ per pixel.}  (Infrared Imager and Spectrometer 2) on
the 3.9m Anglo Australian Telescope (AAT) at Siding Spring
Observatory. The Ks band observations were taken on the 25$^{th}$ of
July 2002 and the J/H band observations carried out on the
16$^{th}$-17$^{th}$ May 2006 with seeing of 1~--~1.2$\arcsec$. For
each source, a 3 $\times$ 3 image grid was created with $\sim$
1$\arcmin$ offsets from the pointing centre. The integration time at
each of the nine grid points was around one minute with 3$\times$20s,
6$\times$10s and 9$\times$6s exposures at J, H and Ks
respectively. The detector bias was removed at readout time using the
Double Read Mode. The data were reduced using the in-house ORAC-DR
pipeline with the `JITTER\_SELF\_FLAT\_KEEPBAD' ORAC-DR\footnote{see
http://www.oracdr.org} recipe to correct for dark current, create
flat-field images from star free pixels in the 9 jittered source
fields and apply a bad pixel mask to each image. The pipeline then
corrects for distortion at the outer edges of the image caused by the
slightly curved focal plane. The 9 individual fields of
7.7$\arcmin$$\times$7.7$\arcmin$ were finally aligned and mosaiced
together to give a 9.7$\arcmin$$\times$9.7$\arcmin$ field of view,
although only the inner 5.7$\arcmin \times$5.7$\arcmin$ are covered by
all 9 fields. The blind telescope pointing error is estimated to be
$\sim$3$\arcsec$ but absolute coordinates were calculated by comparing
non-saturated stars to those in the 2MASS catalogue. Using the
\emph{koords} program in the
KARMA\footnote{http://www.atnf.csiro.au/computing/software/karma/}
visualisation package, the image coordinates were matched to better
than 0.1 pixel accuracy. We take the error in the absolute coordinates
to be the error in the 2MASS catalogue of 0.3$\arcsec$.

The photometry was calibrated from the 2MASS catalogue by plotting the
magnitude difference between matched stars as a function of the
measured IRIS2 magnitude and the mean offset calculated over a
suitable magnitude range ($\sim$12 to 14). The standard deviation in
this range is $\sim$ 0.2 mag but some of this scatter is likely to be
due to the difference between the MKO photometric system used by IRIS2
and the 2MASS photometric system\footnote{See
http://www.astro.caltech.edu/$\sim$jmc/2mass/v3/transformations/}. Although
converting to the 2MASS photometric system may marginally reduce this
scatter, the colour correction terms are not characterised for sources
with (J-H) $>$ 1.5, (H-K) $>$ 1 and (J-K) $>$ 2. As we are primarily
interested in the sources with reddest colours, the photometry has
therefore been left in the MKO system.

Sources were extracted from each of the IRIS2 images using the
\emph{daophot} tasks in the IRAF\footnote{http://iraf.noao.edu/}
package to produce a J, H and Ks star catalogue. Examination of the
residuals after removing the stellar flux from the image shows the
stars have been well extracted despite different background levels
across the image (due to varying extended emission) and the crowded
field. We found 8667, 11762 and 11226 sources at J, H and Ks band
respectively, with a median photometric error from the automated
fitting of $\sim$0.02 mag. The stars in the individual catalogues with
$>$3$\sigma$ detections were then matched as outlined in
$\S$~\ref{subsec:match_catags} to produce a final IRIS2 JHKs
catalogue\footnote{Table~\ref{tab:online_tab} shows an extract from
the catalogue. The full catalogue will be made available on-line.}.

\begin{figure*}
  \begin{center}
    \includegraphics[width=0.3\textwidth, angle=0, trim=0 0 -5 0]{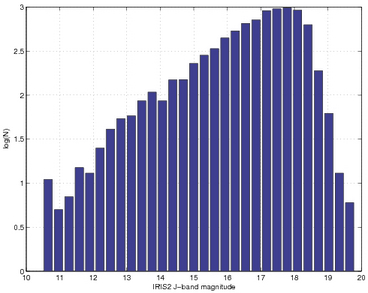}
    \includegraphics[width=0.3\textwidth, angle=0, trim=0 0 -5 0]{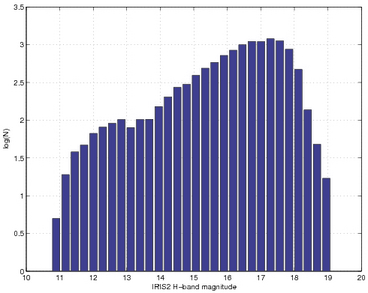}
    \includegraphics[width=0.3\textwidth, angle=0, trim=0 0 -5 0]{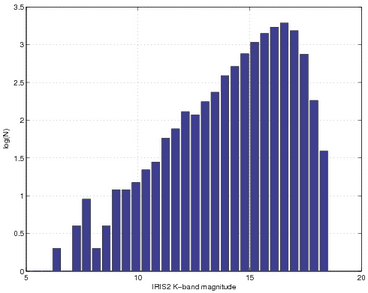}
    \caption{Histograms of the J (left), H (centre) and Ks (right)
    star magnitudes extracted from the IRIS2 images. The 90\%
    completeness limit from artificial star recovery and maximum
    measured magnitudes are 18 \& 19.8, 17.5 \& 19.1, 16.5 \& 18.5 at
    J, H and Ks respectively.}
    \label{fig:jhk_mag_hists}
  \end{center}
\end{figure*}

\subsection{Previous IRIS2 observations}
The previous IRIS2 observations carried out by \citet{leistra2005}
were also made using the J, H and Ks band filters with a similar
integration time and observation method but were instead focused on
the cluster centred at $\alpha_{J2000}$~=~13:11:39.4,
$\delta_{J2000}$~=~-62:33:11. R~$\simeq$~2300 spectra with a
1$\arcsec$-wide slit were also taken toward the cluster. Analysis of
the spectra and photometry revealed G305.24+0.204 as an OB association
with an O5-O6V star and extinction of A$_v\sim$12 mags.

\section{Catalogue Data}
\subsection{2MASS - 2 micron All Sky Survey}
Using two 1.3m telescopes in Arizona and Chile, the 2MASS survey
scanned the entire sky at the J, H and Ks bands (1.25, 1.65 and
2.17$\mu$m) \citep{skrutskie2006}. The survey has a limiting magnitude
of 15.8, 15.1, 14.3 $\pm$ 0.03 mag at J, H and Ks respectively with a
pixel size of 2.0$\arcsec$ and 0.1$\arcsec$ pointing accuracy.

\subsection{GLIMPSE - Galactic Legacy Infrared Mid-Plane Survey Extraordinaire}
\label{sub:glimpse}
Using the infrared array camera (IRAC) on-board the Spitzer Space
Telescope (SST\footnote{http://ssc.spitzer.caltech.edu/}), the GLIMPSE
survey has observed the galaxy at 3.6, 4.5, 5.8 and 8.0$\mu$m with a
1.2" pixel size between $10^\circ<|l|<65^\circ$ and $|b|<1^\circ$
\citep{benjamin2003}.  The photometric accuracy is 0.2 and 0.3 mag for
bands 1/2 and 3/4 respectively. The astrometric accuracy of the point
source catalogue is $\sim$0.3$\arcsec$.

\section{The Point Source Catalogue}
\label{sec:psc}
\subsection{Matching sources at different wavelengths}
\label{subsec:match_catags}
Analysis of source colours relies on the ability to accurately match
sources in the catalogues at different wavelengths. With too many sources
($>$10$^4$) to check manually and crowded stellar fields, there is a
potential problem of either matching a source multiple times or
mismatching the sources completely. The resulting mismatched sources may
replicate the large colour differences of embedded sources and hence
contaminate the sample. To test the matching accuracy, we generated
and matched synthetic catalogues with known offsets from a list of
absolute source positions. In order to ensure a realistic spatial
distribution of sources, including over densities due to clustering
and under densities due to extinction, the absolute coordinates were
taken from the 2MASS PSC of the region. The coordinates recorded in
the synthetic catalogues were randomly offset from the absolute values
by a 2D Gaussian distribution. Several synthetic catalogues were
produced using the known absolute pointing error of the relevant
datasets to generate the positional offsets. The sources in the
synthetic catalogues were given a unique ID and then matched in the
same way as the observed datasets using the \emph{tmatch} routine in
the \emph{tables.ttools} package in IRAF.  The routine matches every
source in the first catalogue with the nearest counterpart in the
second catalogue within a user defined matching radius. Comparing the ID
of the matched sources in the synthetic catalogues it is possible to
unambiguously identify which sources have been matched correctly.

The IRIS2 images were all registered with the same 2MASS image so have
a \emph{relative} astrometric uncertainty of
$\Delta_{IRIS2}=0.04\arcsec$ (0.1 pixel). The astrometric uncertainty
in the 2MASS and GLIMPSE surveys are both quoted as
0.3$\arcsec$. However, the GLIMPSE survey coordinates were also
registered with 2MASS data so the relative uncertainty between the
IRIS2 and GLIMPSE catalogues should be considerably less than the
assumed error of $\Delta_{IRIS2-GLIMPSE}=0.3\arcsec$.  Matching two
synthetic catalogues generated with an uncertainty of
$\Delta_{IRIS2}$, an equal number of sources and a search radius of 2
pixels, produces a 100\% recovery with no confusion due to doubles or
mismatches. Matching catalogues generated with an uncertainty of
$\Delta_{IRIS2}$ but different number of sources also recovers 100\%
of the sources correctly but includes $\sim$0.05\% double
sources. With different numbers of sources in each catalogue, the
matching order then becomes important. For example, matching the
catalogue with larger to smaller number of sources may produce
`multiple matches', where several sources in the larger catalogue are
within the matching radius of a source in the smaller
catalogue. However, the number of correctly matched sources is the
same, irrespective of the matching order. Varying the matching radius
has little effect as long as it is significantly larger than the
combined astrometric uncertainty. A matching radius of 1.2$\arcsec$
was used throughout. We concluded that mismatches do not seriously
affect our results.

The IRIS2 PSC was generated by matching the sources found at each
wavelength in turn from shortest to longest wavelength. To ensure no
red sources were missed, the non-matched sources at the longer
wavelength were also appended to the catalogue after each step. The
2MASS PSC of the same region was used to test the colour/magnitudes of
the brighter IRIS2 PSC sources. The IRIS2 PSC was then matched with
the GLIMPSE PSC to produce the final combined IRIS2/GLIMPSE PSC. A
2MASS/GLIMPSE PSC of the region was also created in a similar way as a
comparison. Using this method, we expect to recover at least 99.8\% of
the sources correctly with a maximum of 0.1\% doubles and 0.12\%
mismatches.

\subsection{Completeness}
\label{subsec:completeness}
We used artificial star recovery to investigate the spatial variation
in point source senstivity as a function of wavelength. Having
calculated the PSF for each image, we inserted a grid of artificial
stars of the same magnitude seperated by 30$\arcsec$ across the
image. We then used the same automated finding technique outlined in
$\S$~\ref{sec:iris2_obs} to calculate how many of the artificial stars
were recovered. By shifting the 30$\arcsec\times$30$\arcsec$ grid of
artificial stars in small steps through the image, it was possible to
measure the completeness at 5$\arcsec$ intervals without the PSF of
individual artificial stars overlapping. The process was repeated by
increasing the artificial star magnitudes in steps of 0.5 mag until no
more stars were recovered. By recording the largest recovered
magnitude at each position and wavelength it was possible to build a
three-dimensional picture of the point source sensitivity across the
region. The \emph{relative} completeness as a function of position was
then calculated at each wavelength by subtracting the median
completness magnitude at that wavelength from every position. This
method is similar to that used by \citet{gutermuth2005} in an analysis
of completeness limits of Spitzer IRAC data.

\begin{figure*} 
  \begin{center} 
    \begin{tabular}{cc} 
\includegraphics[width=7.5cm, height=7.5cm, angle=0, trim=0 0 -5 0]{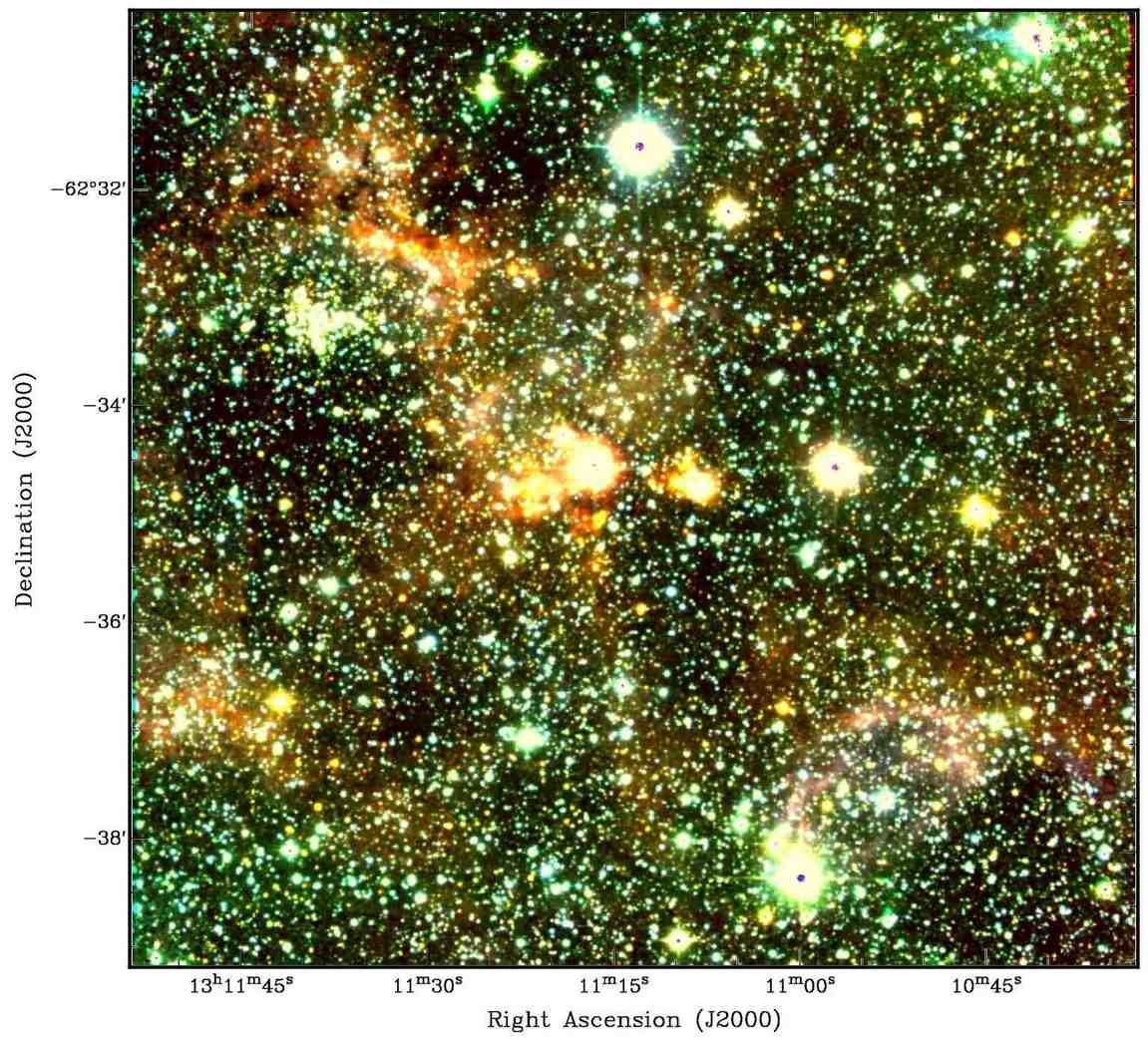} & 
\includegraphics[width=8.0cm,  height=8.0cm, angle=0, trim=0 0 -5 0]{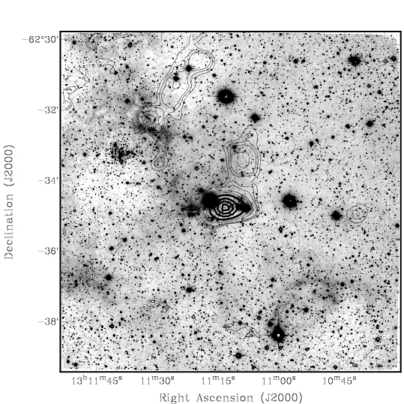} \\ 
    \end{tabular}
  \end{center}
  \caption{a (Left) Three colour (J, H and Ks) IRIS2 image of the
  region G305.2+0.2. b (Right) The IRIS2 Ks-band image, overlayed with
  contours of 1.2mm continuum dust emission from \citet{hill2005}.}
  \label{fig:3col_iris2} 
\end{figure*}

\begin{figure*}
  \begin{center} 
    \begin{tabular}{ccc} 
\includegraphics[width=5.6cm, angle=0, trim=0 0 -5 0]{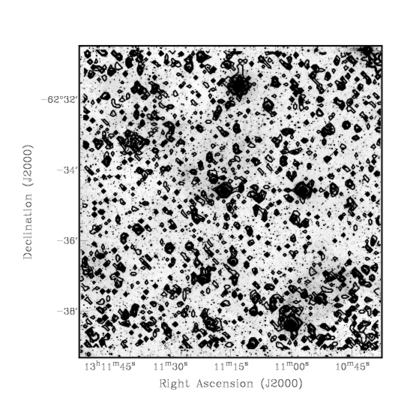} & 
\includegraphics[width=5.6cm, angle=0, trim=0 0 -5 0]{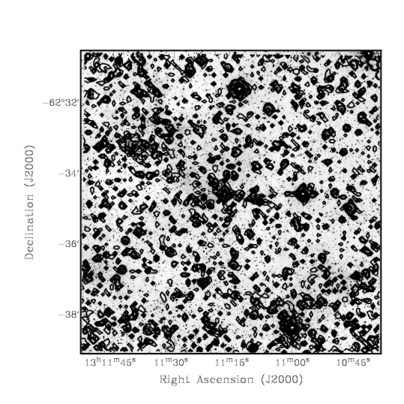} &
\includegraphics[width=5.6cm, angle=0, trim=0 0 -5 0]{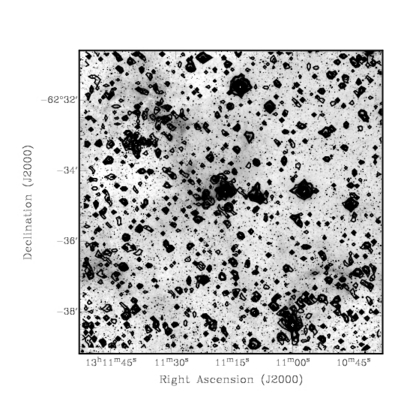} \\ 
    \end{tabular}
  \end{center}
  \caption{The spatial variation in completeness for the J (left), H
(centre) and Ks (right) images calculated as outlined in
$\S$~\ref{subsec:completeness} overlayed on the images at the same
wavelengths. The contours show the relative completeness at -0.5,
-1.0, -1.5 and -2.0 mags, highlighting the areas with poorest point
source sensitivity. An online version in colour illustrates the
detail.}
  \label{fig:jhk_comp} 
\end{figure*}

Figure~\ref{fig:jhk_comp} illustrates the spatial variation in
completeness for the J, H and Ks images. The contours show the
relative completeness at $-$0.5, $-$1.0, $-$1.5 and $-$2.0 mags
highlighting the areas with poorest point source sensitivity. Compared
to the three colour image in Figure~\ref{fig:3col_iris2}a it is clear
that although there is some extended emission, particularly at Ks, the
main factors contributing to decreased sensitivity are saturated
stars, stellar crowding (e.g. the cluster toward the North-East) and
the lower integration time at the edge of the images due to
mosaicing. Figure~\ref{fig:glimpse_comp} shows similar contours at 3.6
and 4.5$\mu$m over the IRIS2 field of view overlayed on the GLIMPSE
images. This time, extended emission is the dominant factor
contributing to decreased completeness.

The 90\% completeness limit calculated from the artificial star
recovery methods above is 18, 17.5 and 16.5 at J, H and Ks
respectively. Figure~\ref{fig:jhk_mag_hists} shows a histogram of the
calculated J, H and Ks magnitudes. The turnover in the number of stars
per magnitude and faintest measured magnitudes are 17.8 \& 19.8, 17.3
\& 19.1, 16.6 \& 18.5 at J, H and Ks respectively. These limiting
magnitudes are fainter than those reported by \citet{leistra2005} of
16, 18 and 18.5 due to confusion towards the cluster, G305.24+0.204,
which was the focus of their observations.

We also note that the turnover is nearly the same as the 90\%
completeness limit found from the artificial star recovery
method. Since it is considerably easier to determine the turnover, it
provides a quicker way of estimating the completeness limit.

\subsection{Control Field}
\label{subsec:control}
\begin{figure}
  \begin{center} 
  \includegraphics[width=5.5cm, angle=-90, trim=0 0 -5 0]{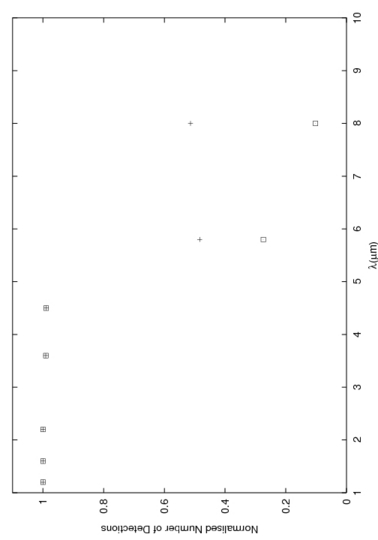}\\ 
  \end{center} 
    \caption{Number of sources detected at each wavelength in the
    combined 2MASS and GLIMPSE catalogues of the target (square) and
    control fields (plus symbols) as outlined in
    $\S$~\ref{subsec:control}.  The number of sources at each
    wavelength are normalised to those detected at 1.2$\mu$m. }
  \label{fig:cont_filters} 
\end{figure}

In order to compare the sources in the target region with a field star
population, a nearby region with no extended emission in either the
GLIMPSE or 2MASS images (centred at $\alpha_{J2000}$~=~12:54:10.3
$\delta_{J2000}$~=~-62:16:12, \emph{l}=303.25 \emph{b}=+0.6) was
chosen as a control field. All sources in the GLIMPSE and 2MASS
catalogues within a 9.7$\arcmin \times$9.7$\arcmin$ box centred on
this position were matched using the same procedure outlined
previously. Figure~\ref{fig:cont_filters} shows the number of sources
detected at each wavelength normalised to the number of sources
detected at 1.2$\mu$m for the target (square) and control fields (plus
symbols). In both fields, the number of sources matched up to
4.5$\mu$m is similar with a sharp drop at longer wavelengths, as most
field stars are too faint to detect. This is exacerbated in the target
field by the extended emission at 5.8 and 8.0$\mu$m.
In $\S$~\ref{subsub:ir_excess} the source colours derived from
2MASS/GLIMPSE for the target and control regions are compared to those
derived from the IRIS2/GLIMPSE PSC.

\subsection{Nature of the IR excess sources}
\label{subsec:contamination}
We aim to use the catalogue to pick out young stellar objects with
excess IR emission through their derived colours. However, these
highly reddened colours are not unique to young stellar objects. We
are therefore interested in calculating the likely contamination from
other astronomical sources with similar colours -- primarily evolved
stars and background galaxies.

Recent Spitzer observations toward the LMC show that background
galaxies with similar colours to young stellar objects
([3.6]-[8.0]$>$1.5) have the potential to contaminate deep
observations for red objects \citep{blum2006}. It is possible to
calculate this likely contamination using the relative number counts
of foreground stars to background galaxies as a function of observed
magnitude for the IRAC bands reported by \citet{fazio2004}. The galaxy
contribution to the source number counts over this wavelength range
only becomes significant for magnitudes greater than 14. Given the
90\% completeness limit of $\sim$14, 13.5, 11.5 and 11 mags at 3.6,
4.5, 5.8, 8.0$\mu$m, the background galaxies are too faint to cause
contamination.

With luminosities of 10$^3$-10$^4$L$_\odot$, radiative transfer
modelling shows that any asymptotic giant branch (AGB) stars along the
line of sight to the edge of the Galaxy will be detected in the IRAC
bands \citep{groenewegen2006}. The same models show that the derived
colours, which depend on the mass loss rate of each AGB star, cover
the expected colour range for young stellar objects. From photometry
alone, it is therefore not trivial to distinguish between AGB stars
and young stellar objects. However, with a model distribution of AGB
stars in the Galaxy it is possible to predict the expected number of
evolved stars within a given field of view. We used the
\citet{jackson2002} model to calculate the number density of AGB stars
as a function of Galactic radius and height above the Galactic plane,
$|z|$, along our line of sight. By integrating the product of the
number density function with discreet volume elements to the edge of
the Galaxy, we then derived the expected number of AGB stars along the
line of sight. Assuming a Galactic radius of 15kpc, a distance of
8.5kpc to the Galactic centre and using $|z|$=0 (i.e. looking directly
through the Galactic plane) we expect to find $\sim$0.7 AGB stars
along the line of sight. It is therefore unlikely that this region
will suffer from significant evolved star contamination.

\section{Results}
\label{sec:results}
\begin{figure*}
  \begin{center} 
    \begin{tabular}{cc} 
\includegraphics[width=8.5cm, angle=0, trim=0 0 -5 0]{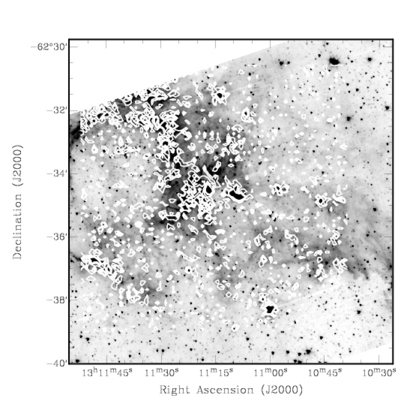} & 
\includegraphics[width=8.5cm, angle=0, trim=0 0 -5 0]{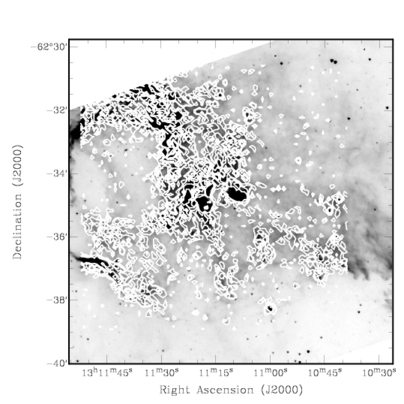} \\ 
    \end{tabular}
  \end{center}
  \caption{The spatial variation in completeness (see
  $\S$~\ref{subsec:completeness}) over the IRIS2 field of view at 3.6
  (left) and 4.5$\mu$m (right) overlayed on the 3.6$\mu$m and
  4.5$\mu$m GLIMPSE images. The contours show the relative
  completeness at -0.5, -1.0, -1.5 and -2.0 mags, highlighting the
  areas with poorest point source sensitivity.}
  \label{fig:glimpse_comp} 
\end{figure*}

Figure~\ref{fig:3col_iris2}a shows an IRIS2, 3-colour image of the
region at near-IR (J, H and Ks) wavelengths and
Figure~\ref{fig:3col_iris2}b shows the Ks image overlayed with 1.2mm
dust continuum emission from \citet{hill2005} to show the most
embedded NIR sources. The image shows lanes of bright extended
emission at 2.2$\mu$m which are seen as red in the 3-colour
image. Figure~\ref{fig:glimpse_comp} shows the GLIMPSE images of the
region at 3.6 and 4.5$\mu$m. The extended emission seen at Ks
dominates the 5.8 and 8.0 $\mu$m GLIMPSE images, making point source
extraction of all but the brightest sources impossible at these
wavelengths. Although the extended emission is seen at 3.6 and
4.5$\mu$m, the images are not affected to the same extent. The nature
of the extended emission is consistent with C-C and C-H bending modes
of polycyclic aromatic hydrocarbon (PAH) emission at 3.3, 6.2, 7.7 and
8.6$\mu$m within the bands excited by UV radiation
\citep{vandishoek2004}. As a result of being unable to extract sources
at the longest wavelength towards the previously studied star
formation regions we may be missing a population of the most embedded
sources.

We find the majority of stars extracted from the catalogues are spread
uniformly across the field, except for the region with the 1.2mm dust
emission and the extended emission at longer IR wavelengths.  The
clear exception is the cluster at $\alpha_{J2000}$=13:11:39,
$\delta_{J2000}$=-62:33:12, with radius $\sim$45$\arcsec$, which abuts
the northern dust filament. The cluster (G305.24+0.204) has previously
been reported by \citet{dutra2003} as an IR cluster and since been
studied in depth by \citet{leistra2005}. We discuss the nature of the
cluster and its relation to the surrounding region in section
$\S$~\ref{sub:cluster_properties}.

\begin{figure*} 
  \begin{center} 
    \begin{tabular}{cc} 
\includegraphics[width=7.5cm, height=7.5cm, angle=0, trim=0 0 -5 0]{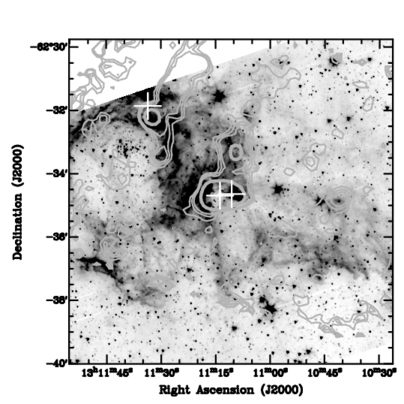} & 
\includegraphics[width=7.5cm,  height=7.5cm, angle=0, trim=0 0 -5 0]{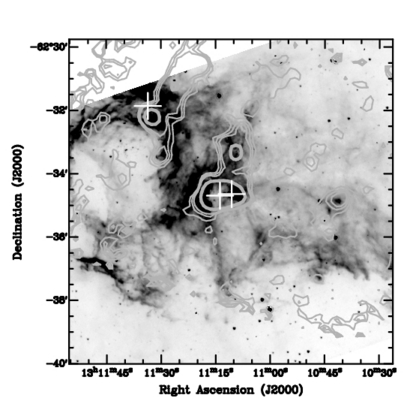} \\ 
    \end{tabular}
  \end{center}
  \caption{GLIMPSE 3.6$\mu$m (left) and 4.5$\mu$m (right) images of
  the region overlayed with the 5, 10, 15 and 50$\sigma$ 1.2mm
  continuum contour levels. The crosses show the position of the
  methanol maser emission.}
  \label{fig:glimpse_images} 
\end{figure*} 

\subsection{Stellar populations}
\label{sub:stellar_pop}
\begin{figure*} 
  \begin{center} 
    \begin{tabular}{cc} 
      \includegraphics[width=7.5cm, height=7.5cm, angle=0, trim=0 0 -5 0]{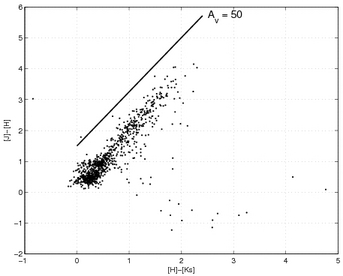} & 
      \includegraphics[width=7.5cm,  height=7.5cm, angle=0, trim=0 0 -5 0]{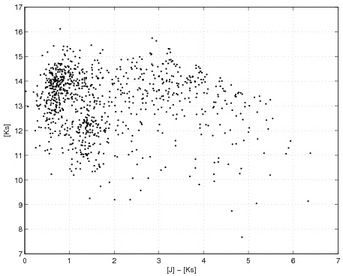} \\ 
    \end{tabular}
  \end{center}
  \caption{(Left) J, H, Ks colour-colour diagram of all stars in the
  IRIS2/GLIMPSE PSC matched in the IRIS2 J, H and Ks catalogues. The
  line gives the reddening vector with a visual extinction of 50
  mags. Inspection of the sources with spurious colours ([J] - [H] $<$
  0, [H] - [Ks] $>$ 0) have been mismatched by the automated matching
  routine ($\S$\ref{sub:stellar_pop}). The number of mismatched
  sources is consistent with that predicted in
  $\S$\ref{subsec:match_catags}.  Very few IR excess sources are
  apparent at these wavelengths. (Right) J, Ks colour-magnitude
  diagram of all stars matched in the IRIS2 J, H and Ks
  catalogues. Although sources are clustered at both $\sim$(0.7, 14)
  and $\sim$(1.4, 12) in the CM diagram, the spatial positions of
  these sources are evenly spread across the field. }
  \label{fig:iris2_ccm_ccd} 
\end{figure*}

To identify the nature of the sources, we analysed the measured star
colours and magnitudes. Figure~\ref{fig:iris2_ccm_ccd} shows
colour-colour (CC) and colour-magnitude (CM) diagrams of all stars in
the IRIS2/GLIMPSE PSC matched in the IRIS2 bands. The [J]-[H] vs
[H]-[Ks] diagram shows that, while the bulk of the sources are
foreground stars, a large fraction are significantly reddened. There
are a small number of sources ($\sim$15) with apparently spurious
colours (e.g. positive [J] - [H], negative [H] - [Ks]). This is
consistent with the predicted number of mismatches in
$\S$~\ref{subsec:match_catags}. Inspection of their SED's reveals two
distinct components: a steeply decreasing flux density at J and H band
followed by an almost constant flux at Ks and above, as expected from
mismatched sources. Visual inspection confirms this result. Apart from
the possible mis-matched sources, there are very few stars with
significant IR excess found using the J, H and Ks colours. The
[J]-[Ks] vs [Ks] diagram shows sources are clustered at both
$\sim$(0.7, 14) and $\sim$(1.4, 12).  However, the spatial positions
of the sources clustered at these points in the CM diagram are evenly
spread across the field.

To ensure these features were real (and as a further check on the
automated star finding/photometry/matching processes), we repeated the
above analysis using the 2MASS only and IRIS2 only PSCs. Both show the
same features with spurious colours in the CC diagram and the clumping
in the CM diagram.

\begin{figure*} 
  \begin{center} 
    \begin{tabular}{cc} 
      \includegraphics[width=7.5cm, height=7.5cm, angle=0, trim=0 0 -5 0]{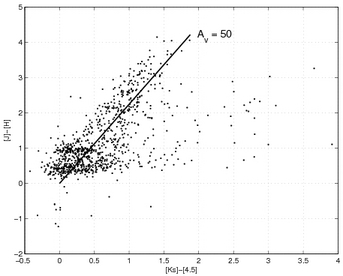} & 
      \includegraphics[width=7.5cm,  height=7.5cm, angle=0, trim=0 0 -5 0]{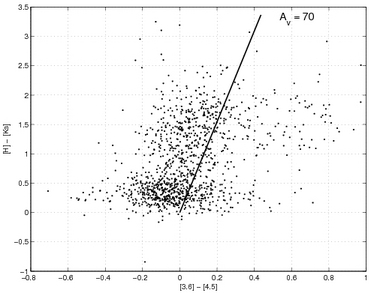} \\ 
    \end{tabular}
  \end{center}
  \caption{(Left) [J] - [H] vs [Ks] - [4.5] and (Right) [H] - [Ks] vs
  [3.6] - [4.5] colour-colour diagrams of stars matched in both the
  IRIS2 and GLIMPSE catalogues. The straight lines show the reddening
  vectors due to dust extinction, corresponding to A$_v$ = 50 and 70
  mags respectively.}
  \label{fig:all_ccm_ccd} 
\end{figure*}

Figure~\ref{fig:all_ccm_ccd} shows selected CC diagrams including the
3.6 and 4.5$\mu$m sources. The combination of longer wavelengths and
larger wavelength baselines in the [J]-[H] vs [Ks]-[4.5] CC diagram
pulls out many more sources with IR excess, confirming that the JHKs
colour combination is not ideal for this purpose. Although the
wavelength baselines are smaller, each of the colours in the [H]-[Ks]
vs [3.6]-[4.5] diagram are generated by measurements on the same
instruments so instrument specific systematic offsets should be
minimised. As in Figure~\ref{fig:iris2_ccm_ccd} there are a number of
sources with spurious colours. These are again consistent with
mismatched sources.

\subsubsection{IR excess sources}
\label{subsub:ir_excess}
\begin{figure}
  \begin{center}
    \includegraphics[width=0.45\textwidth, angle=0, trim=0 0 -5 0]{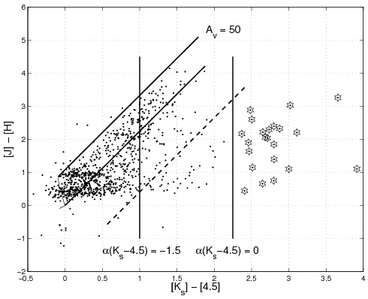}
    \caption{[Ks]-[4.5] vs [J]-[H] colour-colour diagram of stars
      matched in all four bands. Colours for main-sequence and
      super-giant stars are shown as the curved line between
      ([Ks]-[4.5],[J]-[H]) of $\sim$(0,0) and $\sim$(0,1). The two
      solid, non-vertical lines are reddening vectors with A$_v$=50
      mags. The upper line starts from the end of the super-giant
      branch $\sim$(0,1) and the lower line starts from the main
      sequence $\sim$(0,0). The dashed line shows the reddening vector
      with a 3$\sigma$ deviation in the photometric error (see
      $\S$\ref{sec:iris2_obs}) from the main-sequence line (see
      $\S$~\ref{subsub:ir_excess}). All stars under the dashed line
      are classified as having an infrared excess. The two vertical
      lines show the position of stars with spectral indeces,
      $\alpha$, of $-$1.5 (left) and 0 (right). Sources marked with
      stars are those with $\alpha >0$.}
    \label{fig:jhmk_cc}
  \end{center}
\end{figure}

To identify the youngest sources we need to distinguish between stars
with true IR excess and contaminating field stars with red colours due
to extinction. We adopt the relationship derived by
\citet{indebetouw2005}, who show that the extinction due to dust
between 1.2 to 8$\mu$m is well fit with a single slope along several
lines of sight within the galaxy. Figure~\ref{fig:jhmk_cc} shows the
[J]~-~[H] vs [Ks]~-~[4.5] CC diagram of stars matched in all four
bands with two solid, non-vertical lines giving the reddening vectors
of length A$_v$=50 mags. The field star population in the infrared is
likely to be comprised mainly of giants but for completeness, colours
for main-sequence and super-giant stars are also shown as the curved
line between ([Ks]-[4.5],[J]-[H]) of $\sim$(0,0) and $\sim$(0,1) and
the two reddening vectors start from the extremes of the
sequences. Most of the stars lie within or close to these vectors and
are therefore reddened rather than IR excess sources. As the relative
photometric error increases with the measured magnitude, identifying
sources with statistically significant IR excesses will depend on the
source magnitude. However, applying magnitude limits to remove faint
sources has little effect on the reddened sources in the diagram. We
then use a statistical approach to determine which of the reddened
sources are likely to be field stars as opposed to intrinsically red
sources. Assuming the error in the colours is normally distributed,
the number of field sources outside the reddening vectors should drop
off given by the statistics of the normal distribution. The dashed
line shows the reddening vector with a 3$\sigma$ (0.6 mag) deviation
from the main-sequence line which should encapsulate $\sim$99.7\% of
the field sources. In this way, we can calculate the over-abundance of
reddened sources at a given location in the colour-colour diagram and
the likelihood that they are intrinsically reddened. We adopt a value
of 3$\sigma$ of the photometric derived in $\S$\ref{sec:iris2_obs} to
distinguish sources with IR excess and acknowledge a possible
contamination of $\sim$0.3\% from the field source population.

\begin{figure} 
  \begin{center} 
    \includegraphics[width=7.5cm, height=7.5cm, angle=-90, trim=0 0 -5 0]{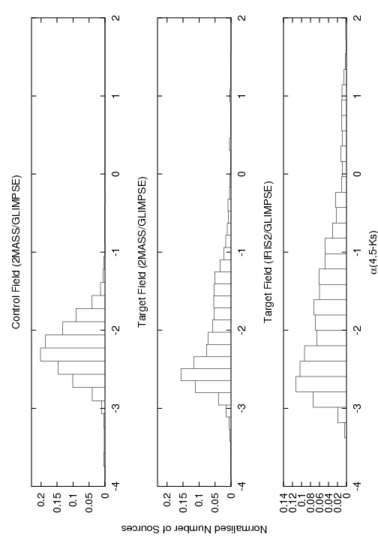}
  \end{center}
  \caption{Histogram of the distribution of spectral indices
  ($\alpha=~d[log(\lambda F_\lambda)]/d[log(\lambda)]$) for all
  sources matched at Ks and 4.5$\mu$m. The top and central figures
  show the distribution in the control field (see
  $\S$\ref{subsec:control}) and target region respectively, for
  sources matched in the 2MASS/GLIMPSE PSC. The bottom figure gives
  the distribution of sources matched in the IRIS2/GLIMPSE PSC of the
  target region.}
  \label{fig:a_k_4.5_hist} 
\end{figure}

Although CC diagrams are powerful tools of finding stars with IR
excess, they can only pick stars matched at all the bands used to make
the plot (i.e. from 1.2 to 4.5$\mu$m). This removes a large number of
potentially highly reddened sources which are not detected at shorter
wavelengths. The spectral index is defined in terms of frequency as,
$n=d[log(\nu F_\nu)]/d[log(\nu)]$ or in terms of wavelength as
$\alpha= d[log(\lambda F_\lambda)]/d[log(\lambda)]$ (where
$n=-\alpha$) and gives the gradient of the IR spectral energy
distribution. As a detection is only required at two frequencies, the
spectral index provides a measure of source reddening which
circumvents the requirement for sources to be observed at all
bands. Figure~\ref{fig:a_k_4.5_hist} shows a histogram of the spectral
index of all sources matched at Ks and 4.5$\mu$m
($\alpha_{Ks-4.5}$). The top and central panels show the distribution
in the control field (see $\S$\ref{subsec:control}) and target region
respectively, for sources matched in the 2MASS/GLIMPSE PSC. Both peak
at $\alpha_{Ks-4.5}\sim~2.5$ but the target region exhibits a strong
tail of red sources compared to the symmetric distribution in the
control region. The bottom panel shows sources matched in the
IRIS2/GLIMPSE PSC of the target region. The distribution is similar to
that found using the 2MASS/GLIMPSE PSC but even more red sources are
detected due to the much deeper near-IR observations.

The spectral index in the near and mid-IR can also be used to infer
the nature of the sources. Normally sources are divided based on the
following criteria: $\alpha <-2$ as Class III, $-2 < \alpha <$ 0 as
Class II, $\alpha >$0 as Class I with Class 0 unobservable in the IR
\citep{adams1987,wilking1989}. However, Figure~\ref{fig:jhmk_cc} shows
that there are a substantial fraction of highly reddened field stars
in this region with spectral indices in this range. To seperate the
highly reddened field stars from the embedded sources we first divide
the full source population into three groups (Group I, II and III)
with the same spectral index cuts used for Class I to III. Using the
[Ks]~-~[4.5] vs [J]~-~[H] CC diagram (Figure~\ref{fig:jhmk_cc}) we can
then investigate the likely field star contamination in each of the
groups. The two vertical lines in Figure~\ref{fig:jhmk_cc} show the
position of stars with spectral indexes of $\alpha = -1.5$ (left) and
0 (right). Clearly, sources with $\alpha < -1.5$ are almost
exclusively field stars. The $\alpha = -1.5$ line provides a much
better cut than $\alpha = -2$ to separate the significant excess
sources and is used from now on to define the cut off between Groups
II and III. The large extinction (A$_v \sim$50 mags) also means only
24\% of the sources with $-1.5< \alpha <$0 are true IR excess sources
-- most are in fact just highly reddened field stars. The final
selection of IR excess sources from the spectral index therefore only
included sources with $\alpha>0$ and those below the 3$\sigma$
photometric error reddening vector.

\begin{figure*} 
  \begin{center} 
    \begin{tabular}{ccc} 
     \includegraphics[width=5.5cm, angle=0, trim=0 0 -5 0]{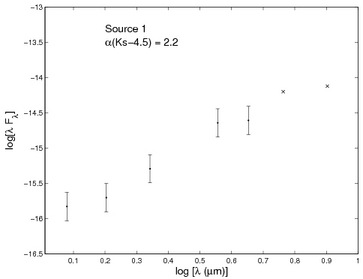} & 
     \includegraphics[width=5.5cm,  angle=0, trim=0 0 -5 0]{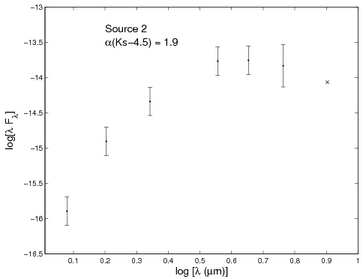} & 
     \includegraphics[width=5.5cm,  angle=0, trim=0 0 -5 0]{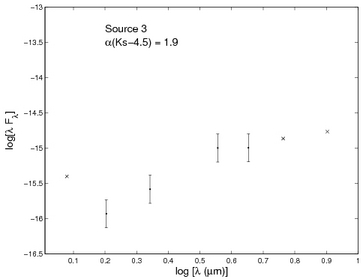}\\ 
      \includegraphics[width=5.5cm,  angle=0, trim=0 0 -5 0]{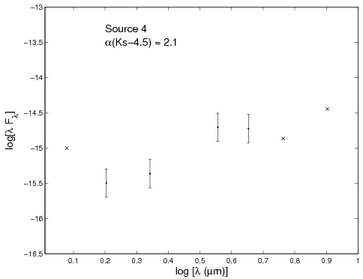} & 
      \includegraphics[width=5.5cm,  angle=0, trim=0 0 -5 0]{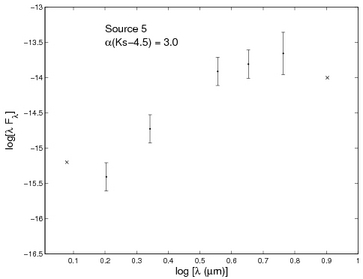} & 
      \includegraphics[width=5.5cm,  angle=0, trim=0 0 -5 0]{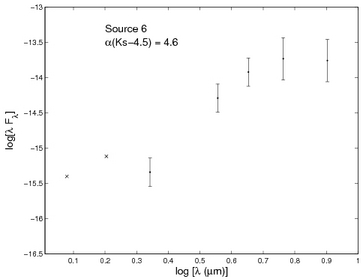}\\ 
      \includegraphics[width=5.5cm,  angle=0, trim=0 0 -5 0]{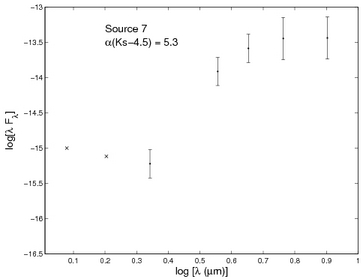} & 
      \includegraphics[width=5.5cm,  angle=0, trim=0 0 -5 0]{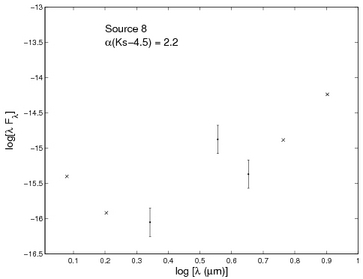} & 
      \includegraphics[width=5.5cm,  angle=0, trim=0 0 -5 0]{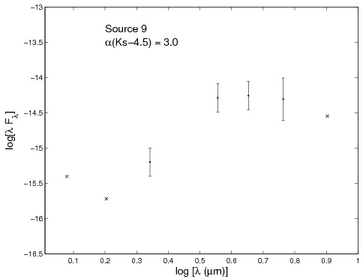}\\ 
      \includegraphics[width=5.5cm,  angle=0, trim=0 0 -5 0]{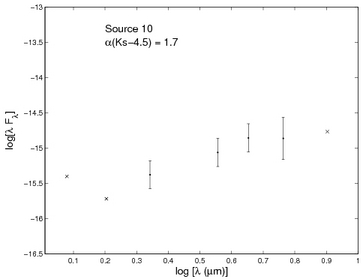} & 
      \includegraphics[width=5.5cm,  angle=0, trim=0 0 -5 0]{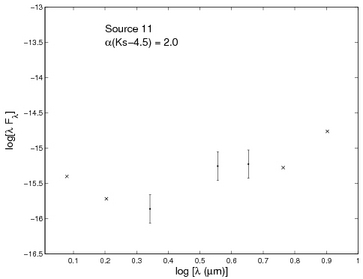} & 
      \includegraphics[width=5.5cm,  angle=0, trim=0 0 -5 0]{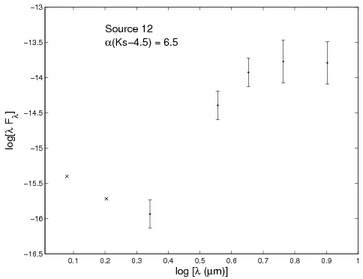}\\ 
    \end{tabular}
  \end{center}
  \caption{Spectral energy distributions for the twelve Group Ie
  sources (defined in $\S$\ref{subsub:ir_excess} as sources with a
  spectral index between Ks and 4.5$\mu$m of greater than 1.5). The
  source number and spectral index calculated between Ks and 4.5$\mu$m
  are given as text for each source. The error bars shown are equal to
  the photometric accuracy of GLIMPSE and the IRIS2 images (see
  $\S$\ref{sec:iris2_obs} \& \ref{sub:glimpse}). For wavelengths with
  non-detections, crosses show the completeness magnitude within the
  nearest $5\arcsec \times 5 \arcsec$ pixel (as described in
  $\S$\ref{subsec:completeness}). It should be noted that these are
  only approximate upper limits due to the fairly coarse (0.5 mag)
  steps used and because the completeness is assumed to remain
  constant over each grid point. With the exception of source 8, the
  calculated spectral index between 2.2 \& 4.5$\mu$m is in good
  agreement with the shape of the SED measured over the complete
  wavelength baseline. }
  \label{fig:irxs_seds} 
\end{figure*}

Table~\ref{tab:num_spect_ind} lists the number of stars in each of the
groups found using spectral indexes of sources detected at all bands
in two different wavelength ranges: the first from J to 4.5$\mu$m
($\alpha_{J-4.5}$) and the second from Ks to 4.5$\mu$m
($\alpha_{K-4.5}$). In total there are 50\% more sources found using
the Ks and 4.5$\mu$m bands. The fraction of Group II and I sources
also increases substantially using the longer wavlengths. There are a
number of Group I sources with extreme IR excess ($\alpha >$ 1.5)
which we define as Group Ie and are potentially even more embedded
than the other Group I objects. Figure~\ref{fig:irxs_seds} plots the
spectral energy distribution (SED) for each of the twelve Group Ie
sources. With the exception of one source (number 8), the spectral
index between Ks and 4.5$\mu$m is a good indication of the SED between
1.2 and 8$\mu$m in general. The fact that most of these 12 sources are
not detected at 5.8 and/or 8.0$\mu$m confirms these would not have
been good filter choices to pick out IR excess sources. Inspecting the
SED's of sources in the other groups results in the same conclusion
that the spectral index between Ks and 4.5$\mu$m is a good indication
of the SED between 1.2 and 8$\mu$m in
general. Figure~\ref{fig:jhmk_cc} suggests it is highly unlikely that
Group I sources are contaminated with field stars and these sources
should therefore be equivalent to Class I embedded sources. However,
care must be taken not to over-interpret the evolutionary stage from
the SED alone as the geometry of these young sources must be, at
minimum, two dimensional \citep{whitney2005}. The measured
colours/spectral indexes are also therefore dependent on the viewing
angle.

\begin{table}
  \begin{tabular}{|c|c|c|c|c|} \hline \hline
    
    Spectral Index     & Group      & N$_{J-4.5}$    & N$_{K-4.5}$ & FS  \\ 
                       &            &                &             & fraction \\\hline
    $\alpha< -1.5$     & III        & 766 (82\%)  & 899 (64\%)     & $\sim$1\\
    $-1.5 <\alpha< 0$  & II         & 147 (16\%)  & 392 (28\%)     & $\sim$0.76\\
    $0<\alpha< 1.5$    & I          & 20  (2\%)   & 99 (7\%)       & $<$0.002\\
    $\alpha> 1.5$      & Ie         & 2   (0.2\%) & 12 (0.9\%)     & -\\ \hline
    \multicolumn{2}{c}{Total}       & 935         & 1402           &\\ \hline \hline

  \end{tabular}
  \caption{Number of stars as a function of spectral index. The group
   names and spectral index selection is explained in
   $\S$~\ref{subsub:ir_excess}. N$_{J-4.5}$ are the number of stars in
   each of the groups which were matched in J, H, Ks and 4.5$\mu$m
   bands. Likewise, N$_{K-4.5}$ are the number of stars in each of the
   groups which were matched in Ks and 4.5$\mu$m bands only. The
   values in parentheses are the fraction of the total number of
   sources in each of the groups. The final column gives the
   approximate fraction of sources in each of the groups which are
   likely to be field stars rather than stars with intrinsic IR excess
   (as calculated in $\S$~\ref{subsub:ir_excess}).}
  \label{tab:num_spect_ind}
\end{table}

The expected lifetime of Class I, II and III sources should be
reflected in the relative number of sources in the different
classes. For low mass stars these ages are around 10$^5$, 10$^6$ and
10$^7$ years for Class I to III respectively
(e.g. \citet{lada1999}). Although Figure~\ref{fig:jkk_clust_cmd} shows
that the detected IR excess sources are likely to be more massive than
this, these ages provide a strong upper limit as massive stars form
much more quickly than their lower mass counterparts.

The significant field star contamination in Groups II and III provides
a strong upper limit to the number of Class II and III
sources. However, without an accurate source count in each Class, an
in-depth analysis is not possible. Assuming the age ratios of the
observed IR excess sources are similar to the low mass case, there
appears to be a large over-abundance of Class I sources (see
Table~\ref{tab:num_spect_ind}). In any case, the much shorter
formation timescales for the more massive stars suggests there has
been a recent epoch of star formation within the region.

\subsubsection{Spatial Distribution of Sources}
\label{subsub:spat_dist}

\begin{figure*} 
  \begin{center} 
    \begin{tabular}{cc} 
      \includegraphics[width=8.0cm,  angle=0, trim=0 0 -5 0]{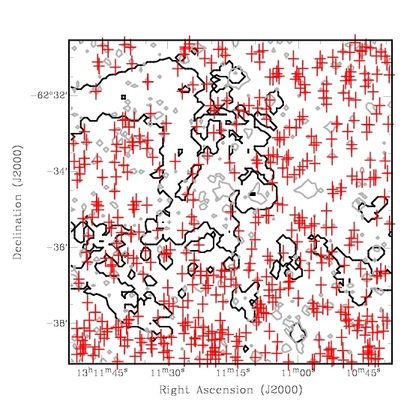} & 
      \includegraphics[width=8.0cm, angle=0, trim=0 0 -5 0]{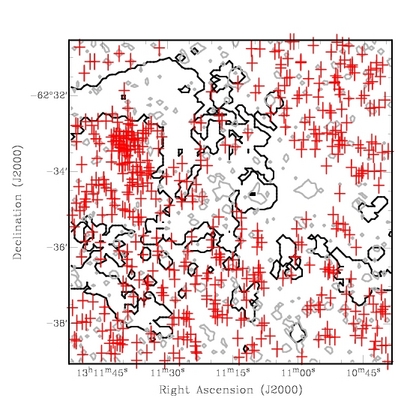} \\ 
      \includegraphics[width=8.0cm, angle=0, trim=0 0 -5 0]{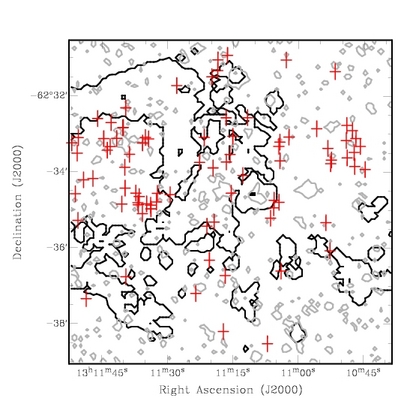} & 
      \includegraphics[width=8.0cm, angle=0, trim=0 0 -5 0]{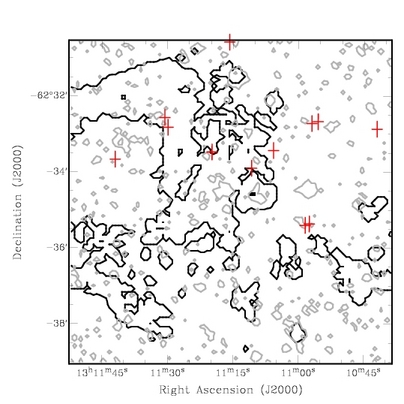} \\ 
    \end{tabular}
  \end{center}
  \vspace{-0.5cm}
  \caption{Source positions (shown as crosses) as a function of the
  spectral index between Ks and 4.5 $\mu$m, $\alpha(Ks-4.5)$. The
  contours show the median subtracted completeness at $-$0.5 mags for
  Ks (grey) and 4.5$\mu$m (black) highlighting areas of poor point
  source sensitivity at each wavelength. For clarity, the 4.5$\mu$m
  completeness image was smoothed using a 15$\arcsec
  \times$15$\arcsec$ kernel before generating the contours. Areas
  outside these contours are termed regions of `moderate
  completeness'. Source positions are separated into their respective
  groups (see $\S$\ref{subsub:ir_excess} for details) as follows:
  Group III (top left), Group II (top right), Group I (bottom left)
  and Group Ie (bottom right). }
  \label{fig:alpha_maps} 
\end{figure*}

Figure~\ref{fig:alpha_maps} shows the distribution of sources (as
crosses) across the region based on their groups as determined from
the spectral index between Ks and 4.5$\mu$m (described in
$\S$\ref{subsub:ir_excess}). The contours show the median subtracted
completeness at $-$0.5 mags for Ks (black) and 4.5$\mu$m (grey)
highlighting areas of poorer point source sensitivity at each
wavelength. For clarity, the 4.5$\mu$m completeness image was smoothed
using a 15$\arcsec \times$15$\arcsec$ kernel before generating the
contours.

Before interpreting the spatial distribution of source colour, we
first consider the effects of completeness as a function of wavelength
across the region. The most obvious effect is that regions with few
crosses correlate with those regions with poorest
completeness. Investigation of the source colours in these regions,
including the current sites of massive star formation (MM4 in
Figure~\ref{fig:g305_simba}), will therefore be strongly biased by the
completeness effects. Instead, we focus on the surrounding region in
which the median subtracted completeness is $>-$~0.5 mags at both Ks
and 4.5$\mu$m. Within this `moderate completeness' region, completion
effects will be small and comparable to the photometric error.

The top left, top right, bottom left and bottom right plots in
Figure~\ref{fig:alpha_maps} give the spatial distribution of the Group
III, II, I and Ie sources, respectively (as defined in
$\S$\ref{subsub:ir_excess}).  The Group III and II sources are fairly
evenly distributed throughout the moderate completeness region. The
clear exception is the density peak in the top right image, revealing
that most of the G305.24+0.204 cluster members are Group II
sources. Confusion due to stellar crowding in the cluster is the cause
of the poorer completeness limit at Ks band in this region.

Interestingly, the Group I and Ie distributions shows a substantial
number of IR excess sources lie offset from the known star formation
regions (MM4 in Figure~\ref{fig:g305_simba}) in the moderate
completeness region. Neither the Group I nor Ie sources are confined
to a single area. The Group I sources in particular appear to be
widely spread, with the exception of the two areas near
$\alpha_{J2000}$=13:11:35, $\delta_{J2000}$=-62:34:48 and 13:10:49,
-62:33:20. Although much fewer in number, the 2MASS/GLIMPSE PSC of the
region confirms these results and shows a similar source distribution.

\begin{figure*} 
  \begin{center} 
    \begin{tabular}{cc} 
      \includegraphics[width=8.0cm, angle=0, trim=0 0 -5 0]{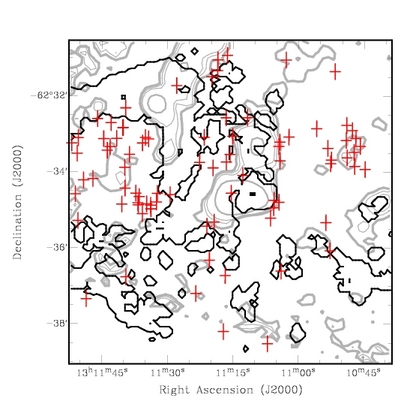} & 
      \includegraphics[width=8.0cm, angle=0, trim=0 0 -5 0]{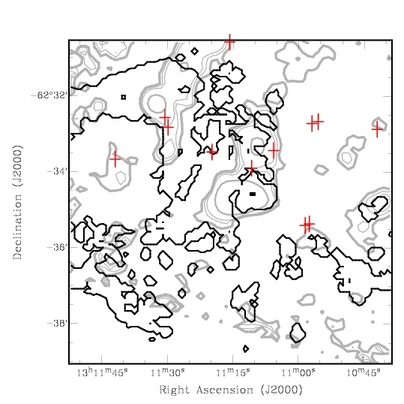} \\ 
    \end{tabular}
  \end{center}
  \vspace{-0.5cm}
  \caption{Source positions (shown as crosses) as a function of the
  spectral index between Ks and 4.5 $\mu$m, $\alpha(Ks-4.5)$, for
  Group I (left) and Group Ie (right) sources. The grey contours show
  the 1.2mm continuum emission at 5, 10, 15 and 50$\sigma$. The black
  contours show the $-$0.5 magnitude median subtracted completeness at
  4.5$\mu$m. For clarity, the 4.5$\mu$m completeness image was
  smoothed using a 15$\arcsec \times$15$\arcsec$ kernel before
  generating the contours.}
  \label{fig:alpha_simba} 
\end{figure*}

Figure~\ref{fig:alpha_simba} shows the position of the Group I and Ie
sources overlayed on the 1.2mm dust continuum emission. Several of the
Group I sources in the IRIS2 field of view lie either completely
outside the 1.2mm field or in regions of much poorer sensitivity
within 60$\arcsec$ of the 1.2mm field edge. As a result, these sources
should be discarded from discussion involving the IR excess source
distribution and the dust distribution. Of the remaining sources in
the moderate completeness region, most are associated with, or lie
close to, dust clumps.

\subsection{Properties of the IR cluster G305.24+0.204}
\label{sub:cluster_properties}
The cluster at $\alpha_{J2000}$=13:11:39.4
$\delta_{J2000}$=-62:33:11.55, designated G305.24+0.204 by
\citet{clarkporter2004} has previously been studied in detail at
near-IR wavelengths by \citet{leistra2005}. With similar observations
to \citet{leistra2005}, we first use the results of their
spectroscopic analysis as a further check on our photometry and
matching algorithms in $\S$~\ref{subsub:compare_cluster}, then discuss
the interaction of the cluster with the surrounding environment in
$\S$~\ref{subsub:environment_cluster} and finally investigate the
embedded stellar populations using the longer wavelength data in
$\S$~\ref{subsub:irxs_cluster}.

\subsubsection{JHKs photometry}
\label{subsub:compare_cluster}
\begin{figure}
  \begin{center}
    \includegraphics[width=0.45\textwidth, angle=0, trim=0 0 -5 0]{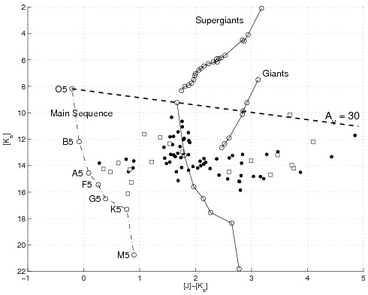}
    \caption{[J]-[Ks] vs [Ks] colour-magnitude diagram of stars
      matched in the two bands within a radius of 45$\arcsec$ from the
      cluster centre ($\alpha_{J2000}$=13:11:39.4,
      $\delta_{J2000}$=-62:33:12), shown as solid dots. The squares
      show stars within the control region, also of radius
      45$\arcsec$. The control and cluster sources have a similar
      distribution with the exception of the clear over abundance of
      sources in the cluster field between [J]-[Ks] of 1.5 to 2.3. The
      dot-dashed line shows the position of the Main Sequence and
      selected stellar spectral types as open circles, with no
      reddening, at a distance of 3.9kpc (the distance to the central
      star forming region). The dashed line shows a reddening vector
      with a length of A$_v$=30 mags. The solid lines show the
      position of the Main Sequence, Giant and Supergiant stars at
      3.9kpc with A$_v$=11~mags with open circles again showing
      selected spectral types. The majority of stars in the cluster
      field (solid circles) lie along this Main Sequence as O or B
      stars. }
    \label{fig:jkk_clust_cmd}
  \end{center}
\end{figure}

To analyse the cluster population we selected all stars within a
radius of 45$\arcsec$ of the cluster centre and a second control
sample of stars within 45$\arcsec$ of the position
$\alpha_{J2000}$=13:11:42.4,
$\delta_{J2000}$=-62:34:32. Figure~\ref{fig:jkk_clust_cmd} shows
the [J]-[Ks] vs [Ks] colour-magnitude diagram of stars matched in the
two bands. The stars near the cluster are shown as dots/triangles and
the control stars as squares. The control stars and the stars in the
cluster field not associated with the cluster are similarly spread
throughout the diagram. The stars associated with the cluster are
clearly concentrated at ([J]-[Ks], [Ks])$\sim$(1.9, 13) which is
consistent with O and B stars at a distance of 3.9kpc with an
extinction of A$_v \sim$11 mags. This is also consistent with previous
distance estimates and calculated extinctions based on spectral
identifications of the most luminous stars in the IR band (see
\citet{leistra2005}). The sources shown as triangles are those within
the cluster region which lie to the red-ward side of the Main Sequence
line defining the cluster members by $>$3$\sigma$ of the photometric
error. These sources either do not belong to the cluster or are
reddened cluster members. IR excess sources toward the cluster are
discussed in $\S$\ref{subsub:irxs_cluster}.

\subsubsection{Interaction with the surrounding environment}
\label{subsub:environment_cluster}
With powerful winds and radiation pressure driven by the OB stars, the
cluster has the potential to interact strongly with the surrounding
environment. Recent observations show there is diffuse 4GHz continuum
associated with and surrounding the cluster (Walsh, private
communication) which appears to be roughly confined by the surrounding
extended emission seen in the infrared. It is plausible that the
cluster of OB stars is therefore responsible for both ionising the
surrounding region (evidenced as cm-continuum emission) and providing
the UV flux exciting the PAH emission which surrounds the cluster (see
Figure~\ref{fig:glimpse_comp}). A reasonable interpretation of this
evidence is that the cluster is driving an expanding shell of
molecular material into the surrounding region, the densest regions of
which are observed as the 1.2mm dust continuum clumps. The fact that
some of these clumps are in the earliest stages of forming stars (as
indicated by the nearby methanol maser and Group Ie sources in the
south-eastern peak of MM3 in Figure~\ref{fig:g305_simba}) suggests the
cluster may have been responsible for triggering a further generation
of star formation.

\subsubsection{IR excess sources}
\label{subsub:irxs_cluster}
\begin{figure*} 
  \begin{center} 
    \begin{tabular}{ccc}
 \includegraphics[width=5.5cm, angle=0]{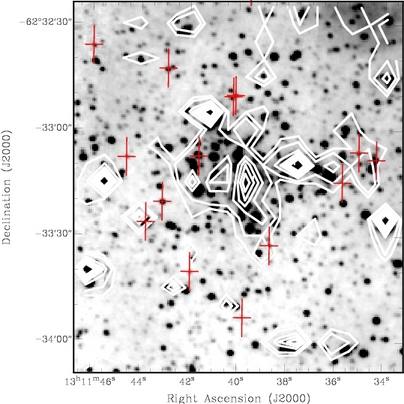}&
 \includegraphics[width=5.5cm, angle=0]{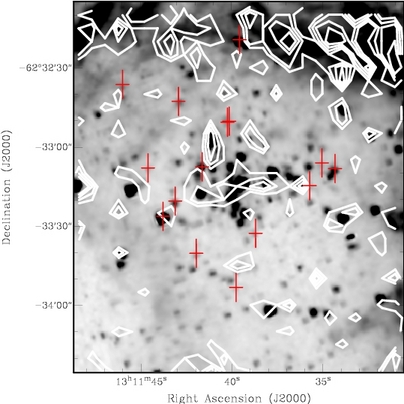}&
 \includegraphics[width=5.5cm, angle=0]{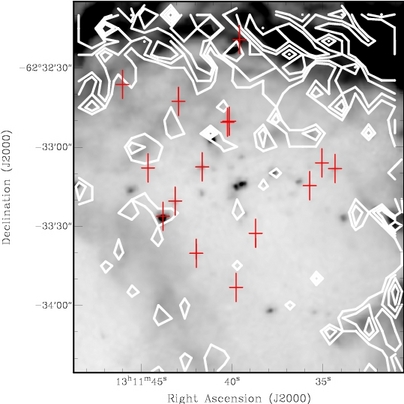}\\
 \end{tabular}
  \end{center}
  \caption{Images at Ks (left), 3.6$\mu$m (centre) and 4.5$\mu$m
  (right) of the IR cluster G305.24+0.204. The contours show the
  median subtracted completeness at each wavelength in steps of $-$0.5
  magnitudes, highlighting areas with poorer point source
  sensitivity. Crosses show the position of IR excess sources
  ($\alpha$(Ks-4.5)~$>$~0, group I and Ie) within 60$\arcsec$ of the
  cluster centre.}
  \label{fig:ak4.5_clust} 
\end{figure*}

\citet{leistra2005} reported no IR excess stars toward the cluster
from J, H and Ks colours. However, including the longer wavelength
data reveals an over density of IR excess sources located around the
cluster (Figure~\ref{fig:ak4.5_clust}). Sources with $\alpha>$~0,
based on the slope of the spectrum from Ks to 4.5$\mu$m, are
distributed around the core of the cluster but none are found at the
centre. From Table~\ref{tab:num_spect_ind} and the arguments in
$\S$~\ref{subsec:contamination}, these are unlikely to be
contaminating reddened field stars or background galaxies. Although
there is no way to unambiguously confirm these sources belong to the
cluster, it seems unlikely that $>$15\% of the IR excess sources in
the entire region would randomly fall within $<$5\% of the total area
(corresponding to a radius of 60$\arcsec$) of a randomly chosen
location. While this is certainly not a rigourous argument (it is
biased by our selection of the radius around the cluster for example)
it provides reasonable justification for assuming that at least some
of these sources are associated with the cluster. In this case, can we
find an explanation for the lack of IR excess objects towards the
cluster centre?

Figure~\ref{fig:ak4.5_clust} shows that confusion due to the high
stellar density towards the cluster centre at Ks leads to
significantly poorer completion limits at this wavelength. It is
therefore possible that by using the spectral index between Ks and
4.5$\mu$m, real IR excess sources detected at 4.5$\mu$m toward the
cluster centre have been missed as they lie beneath the Ks completion
limit. Figure~\ref{fig:ak4.5_clust} illustrates that the completeness
problem drops substantially at 3.6$\mu$m and is almost non existent
toward the cluster centre at 4.5$\mu$m. Therefore, to investigate the
apparent lack of IR excess sources toward the cluster centre, we used
the GLIMPSE only catalogue of the region. This removes any potential
errors due to the automated photometry and matching caused by the high
stellar crowding in the IRIS2 images. In the GLIMPSE PSC, all of the
sources detected at 4.5$\mu$m towards the cluster were also detected
at 3.6$\mu$m, despite the reduced completeness in the 3.6$\mu$m
images. As the 4.5$\mu$m image does not suffer from reduced point
source sensitivity towards the cluster centre (see
Figure~\ref{fig:ak4.5_clust}), we conclude that we have a complete
sample of 4.5$\mu$m sources for the cluster. The spectral index
between 3.6 and 4.5$\mu$m from the GLIMPSE PSC confirms the lack of IR
excess sources toward the cluster centre and again shows IR excess
sources at the cluster edge.

The previous analysis has been based on the major assumption that the
IR excess sources belong to the cluster. Besides their proximity, is
there any reason to believe they may be associated? In comparison to
the Orion Nebula Cluster (ONC), these stars are slightly outside the
radius of $\sim$0.2pc of the stars in the Trapezium which form the
core of the ONC \citep{hillenbrandhartmann1998}. However, they are
well inside the $\sim$2.5pc radius comprising the entire ONC
\citep{hillenbrand1997}, which has in excess of a thousand members
visible in the near-IR. Using the ONC for comparison, it seems
plausible that the IR excess sources could belong to a larger complex
based around the cluster.

\section{Discussion}
\label{sec:discussion}
With an indicator of the source ages and their spatial distribution,
we can investigate the star formation history of the G305.2+0.2
region. Although we cannot probe the most heavily embedded regions due
to incompleteness, the question remains -- how did the detected IR
excess sources form in a short period ($\leq 10^6$ years) over such a
large distance ($\sim$5 pc) and yet are clearly seperated from the
known regions of star-formation activity, such as the UC\hii regions
and maser emission toward MM4? If these are the only sites of recent
star formation, one possible explanation may be that the IR excess
sources have been ejected from them. To investigate the plausibility
of this argument, we consider the angular distance on the sky it is
feasible for a star at a distance of 3.9kpc to have moved within its
lifetime. \citet{hoogerwerf2000,hoogerwerf2001} show that 10\% of
massive stars are high velocity ($>$ 30 kms$^{-1}$) runaways and up to
30\% are moderate velocity (10$<$V$<$30 kms$^{-1}$) runaways. A class
I star with a lifetime of 10$^5$ years and an ejection velocity of
$\sim$10 kms$^{-1}$ could therefore potentially have travelled an
angular distance of 60$\arcsec$ ($\sim$ 1pc) at 3.9kpc.

\citet{huffstahler2006} argue that local potential wells caused by
other stars and ambient molecular gas will stop stars drifting from
the cluster on ballistic trajectories. However, in young, massive,
Trapezium-like systems with stellar densities $\sim$10$^4$pc$^{-3}$
\citep{hillenbrand1997} and large numbers of companions per system,
dynamical ejections are likely.  The potential due to other stars and
even molecular gas is not sufficient to stop the ejection as evidenced
by an ejection event within the last 500 years in the deeply embedded
BN-KL complex in Orion, which produced a runaway OB star
\citep{rodriguez2005}. If the IR excess objects we observed have been
ejected, measurement of their proper motion vectors with
milliarcsecond astrometry over a decade long time scale should reveal
whether they are indeed moving at several kms$^{-1}$ away from the
sites of active star formation. However, while this is certainly
plausible for some sources, particularly those nearest nearest MM4, it
seems unlikely that this can account for all the IR excess
sources. The shorter lifetimes of high mass stars compared to their
lower mass counterparts suggest an unreasonably high velocity for the
sources farthest away from MM4.  There is also no explanation for the
non-homogenous ejections resulting in the two concentrations of IR
excess sources.

Ruling out MM4 as the sole origin of the IR excess sources implies
there must be other sites of recent star formation within the
region. Focusing on regions with moderate completeness limits (see
$\S$\ref{subsub:spat_dist}), the area confined by the PAH emission in
the 3.6 and 4.5$\mu$m images stands out with a high density of IR
excess sources (see Figure~\ref{fig:alpha_maps}). The concentration of
IR excess sources near $\alpha_{J2000}$=13:11:35
$\delta_{J2000}$=-62:34:48 and 13:10:49 -62:33:20 make these strong
candidates for sites of recent star formation. Interestingly, only the
latter of these concentrations is associated directly with a dust core
(MM7). This suggests it has not yet had time to disperse its natal
molecular material and, as such, may be the younger of the two
sites. While IR excess sources are seen towards other dust cores in
the moderate completeness region (MM6, MM8, the north-west of MM3 and
the western side of MM5) the fewer sources numbers make the case for
these regions less clear. Finally, the much poorer sensitivity at the
edge of the 1.2mm field makes it difficult to comment on the
association of IR excess sources with the dust in these regions.

\section{Conclusions}
\label{sec:conclusions}
We have generated a point source catalogue towards the G305.2+0.2
region from 1.2 to 8.0$\mu$m by combining deep, near-IR, IRIS2 images
with catalogued mid-IR data. Modeling of the automated photometric
extraction and star matching algorithms predicts a maximum of 0.12\%
mismatches in the catalogue, which is consistent with the number of
sources with apparently spurious SED's. Through comparison with
previous spectroscopic observations of the region we have confirmed
the accuracy of the photometry. We find:
\begin{itemize}
\item There is strong extended PAH emission in the GLIMPSE
   images. This has the effect of significantly lowering the point
   source sensitivity in these regions, particularly at 5.8 and
   8.0$\mu$m. As a result we may be missing a population of the most
   heavily embedded sources, including those toward the known sites of
   massive star formation.

\item 12 of the sources have extreme IR excess with spectral indices
  between 2.2$\mu$m and 4.5$\mu$m $>$1.5 which we designate Group
  Ie. Analysis of the full SED shows the spectral index over this
  reduced wavelength range is a reliable indicator of the IR SED in
  general.

\item The effect of the winds and radiation from the OB association
  G305.24+0.204 may be responsible for triggering a third generation
  of star formation. 

\item There is a population of IR excess stars surrounding the cluster
   G305.24+0.204 and an absence of such sources in the cluster
   centre. There is reasonable evidence that some of these excess
   sources belong to the cluster. We discuss possible causes for this.

\item There is a sizable population of IR excess sources offset from
  the known star formation sites. Analysis of these sources reveal
  multiple new sites of recent star formation. The formation activity
  in the G305.2+0.2 region is therefore much more widespread than
  previously thought.

\end{itemize}

\section{Acknowledgements}
We would like to thank Stuart Ryder for his help with the data
reduction and Andrew Walsh for discussions of the G305 region. SNL
would like to thank Charlie Lada, Lori Allen, Luisa Rebull and Rob
Gutermuth for help interpretting GLIMPSE data. We thank the anonymous
referee for instructive comments which significantly improved the
clarity of the manuscript. SNL is supported by a scholarship from the
School of Physics at UNSW. This research has made use of NASA's
Astrophysics Data System.

\bibliography{longmore_s_g305}

\input tables/online_tab

\end{document}

%% file: tables/online_tab.tex
\begin{table*}
\begin{scriptsize}
  \begin{center}
  \caption{Example extract from the first five lines of the online
  catalogue of the sources matched between the IRIS2 J, H and Ks band
  images and the GLIMPSE point source catalogue. The first two columns
  give the source right ascension and declination in decimal
  degrees. Columns 3 to 11 give the magnitude, magnitude error and ID
  number of the sources at J, H and Ks band. Columns 12 through 19
  give the magnitude and magnitude error of the sources at 3.6, 4.5,
  5.8 and 8.0\,$\mu$m from the GLIMPSE point source catalogue. The
  final column gives the GLIMPSE ID number. In this example table, the
  magnitudes and errors have been rounded to 2 decimal places for
  clarity. A value of `null' means no source was detected at that
  wavelength.}
  \label{tab:online_tab}

  \begin{tabular}{|c|c|c|c|c|c|c|c|c|c|c|c|c|c|c|c|c|c|c|c|}\hline\hline 
    RA J2000  & Dec. J2000 & J     & $\Delta$J & ID & H     & $\Delta$H & ID & Ks   & $\Delta$Ks & ID & 3.6 & $\Delta$3.6 & 4.5 & $\Delta$4.5 & 5.8 & $\Delta$5.8 & 8.0 & $\Delta$8.0 & ID\\
    (degrees) & (degrees)  & (mag) & (mag)     & (J)& (mag) & (mag)     & (H)&(mag) & (mag)      & Ks &(mag)& (mag)       &(mag)&(mag)        &(mag)&(mag)        &(mag)& (mag)      & GLIMPSE \\ \hline
  197.8149 & -62.65503 & 15.90 & 0.02 & 372 & 14.12 & 0.01 & 522 & 13.27 & 0.01 & 13 & 12.71 & 0.09 & 12.57 & 0.10 & null & null & null & null & 5647 \\
  197.7037 & -62.65393 & 15.36 & 0.02 & 393 & 14.56 & 0.03 & 544 & 14.48 & 0.02 & 33 & 14.08 & 0.14 & 13.81 & 0.14 & null & null & null & null & 449 \\
  197.8973 & -62.65512 & 14.20 & 0.02 & 388 & 13.02 & 0.04 & 542 & 12.36 & 0.03 & 39 & 12.12 & 0.08 & 12.08 & 0.07 & null & null & null & null & 887 \\
  197.916 & -62.65464 & 18.41 & 0.12 & 433 & 16.29 & 0.04 & 596 & 15.16 & 0.03 & 82 & 13.95 & 0.14 & 13.89 & 0.29 & null & null & null & null & 926 \\
  197.7169 & -62.65399 & 17.30 & 0.07 & 403 & 15.14 & 0.05 & 543 & 14.06 & 0.02 & 32 & 13.29 & 0.01 & 13.03 & 0.20 & null & null & null & null & 2111 \\ \hline

  \end{tabular}
  
  \end{center}
  \vspace{-2mm}
\end{scriptsize}
\end{table*}